\documentclass[%
groupedaddress,
preprint,
showpacs,
 amsmath,amssymb,
 aps,
prb,
]{revtex4-1}

\newcommand\beq{\begin{equation}}
\newcommand\eeq{\end{equation}}

\usepackage{graphicx}
\usepackage{dcolumn}
\usepackage{bm}

\begin{document}


\title{Analytical Study of Sub-Wavelength Imaging by Uniaxial Epsilon-Near-Zero Metamaterial Slabs}

\author{Giuseppe Castaldi}
\author{Silvio Savoia}
\author{Vincenzo Galdi}
\email{vgaldi@unisannio.it}
\affiliation{Waves Group, Department of Engineering, University of Sannio, I-82100 Benevento, Italy
}%

\author{Andrea Al\`u}
\affiliation{Department of Electrical and Computer Engineering, The University of Texas at Austin, Austin, TX 78712, USA}%

\author{Nader Engheta}
\affiliation{Department of Electrical and Systems Engineering, University of Pennsylvania, Philadelphia, PA 19104, USA}%

\date{\today}


\begin{abstract}
We discuss the imaging properties of uniaxial epsilon-near-zero metamaterial slabs with possibly tilted optical axis, analyzing their sub-wavelength focusing properties as a function of the design parameters. We derive in closed analytical form the associated two-dimensional Green’s function in terms of special cylindrical functions. For the near-field parameter ranges of interest, we are also able to derive a small-argument approximation in terms of simpler analytical functions. Our results,
validated and calibrated against a full-wave reference solution, expand the analytical tools available
for computationally-efficient and physically-incisive modeling and design of metamaterial-based sub-wavelength imaging systems.
\end{abstract}

\maketitle

\section{Introduction}

The seminal work by Pendry \cite{Pendry} demonstrated that a slab of negative-index material (lossless and impedance-matched with the surrounding medium) would ideally image a perfect copy of a source. Such {\em perfect lensing} mechanism allows the restoration of the sub-wavelength spatial details carried by the evanescent spectral components via efficient coupling to resonant surface-plasmon states, and may lead to revolutionary applications in a variety of fields, including nanolithography, bio-sensing, and spectroscopy. At infrared and optical frequencies, where magnetic activity is difficult to achieve and many materials may naturally exhibit negative permittivity, an approximate implementation was suggested \cite{Pendry} and experimentally demonstrated \cite{Liu,Melville,Fang} in the simple form of a thin layer of silver, with severe restrictions on the field polarization and near-field range, and the sub-wavelength resolution ultimately limited by the material losses.

The above results have generated a growing interest in the study of metallo-dielectric multilayered structures (see, e.g., Refs. \onlinecite{Ramakrishna1, Ramakrishna2, Cai, Wood, Yang, Salandrino, Belov2, Jin1, Belov3, Luo, Stefaniuk, Jin2, Pastuszczak, Liu1, Pastuszczak} for a sparse sampling), which allow mitigation of the above loss- and range-related limitations, as well as further degrees of freedom for design optimization. Essentially, these configurations exploit the inherent (possibly {\em extreme}) anisotropy exhibited by metallo-dielectric multilayers in order to convert evanescent spectral components with large transverse wavenumbers into propagating waves. More recently, the use of {\em obliquely} layered structures has been proposed in order to achieve simple image manipulation (lateral displacement) with sub-wavelength resolution. \cite{Wang} 
Of particular interest is the {\em epsilon-near-zero} (ENZ) regime, which has also found interesting applications to other scenarios, including cloaking,\cite{ENZ5,ENZ6}
light {\em funneling} through sub-wavelength apertures,\cite{ENZ1} 
controlled leaky-wave radiation,\cite{ENZ2} suppression of Anderson localization in disordered multilayers,\cite{ENZ3} 
loss-induced omnidirectional bending,\cite{ENZ4}  
and nonlinearity enhancement. \cite{ENZ7,ENZ8}  
Also worth of mention are the studies on sub-wavelength imaging devices based on wire media (see, e.g., Refs. \onlinecite{Belov1,Belov6,Belov7}), as well as on the transformation-optics paradigm (see, e.g., Refs. \onlinecite{Tsang,Yan,Gallina}). 

With the exception of few cases (see, e.g., Refs. \onlinecite{Wood,Liu1}), for which analytical approximations of the Green's function can be worked out, the imaging properties (e.g., resolution) of the above configurations need to be assessed numerically.
In this paper, we study in more detail this geometry and we show 
that the two-dimensional (2-D) Green's function of a slab of uniaxial ENZ metamaterial with tilted optical axis can be calculated analytically in closed-form, greatly facilitating the analysis and design of its anomalous imaging properties.
 Our results are expressed in terms of special cylindrical functions that can be efficiently computed, and are also amenable to simple approximations in the parameter range of interest.  These findings shed new light on the sub-wavelength imaging properties of ENZ metamaterial slabs and allow tailoring their properties without the need of extensive numerical simulations.

The rest of the paper is organized as follows. In Sec. \ref{Sec:Geom}, we introduce the problem geometry and formulation. In Sec. \ref{Sec:Analytic}, we derive the analytical solution for the Green's function (with the more involved details relegated in Appendices \ref{AppA}--\ref{AppD}), discuss the related computational issues, and work out more manageable approximations
for specific values of the design parameters of interest. In Sec. \ref{Sec:Results}, we illustrate some representative results in order to validate and calibrate our proposed solutions,  and discuss relevant physical insights on the imaging properties of the considered metamaterial lenses.
Finally, in Sec. \ref{Sec:Conclusions}, we provide some brief conclusions.

\section{Problem Geometry and Formulation}
\label{Sec:Geom}
\subsection{Generalities}
\label{Sec:Gen}
\begin{figure}
\begin{center}
\includegraphics [width=8.5cm]{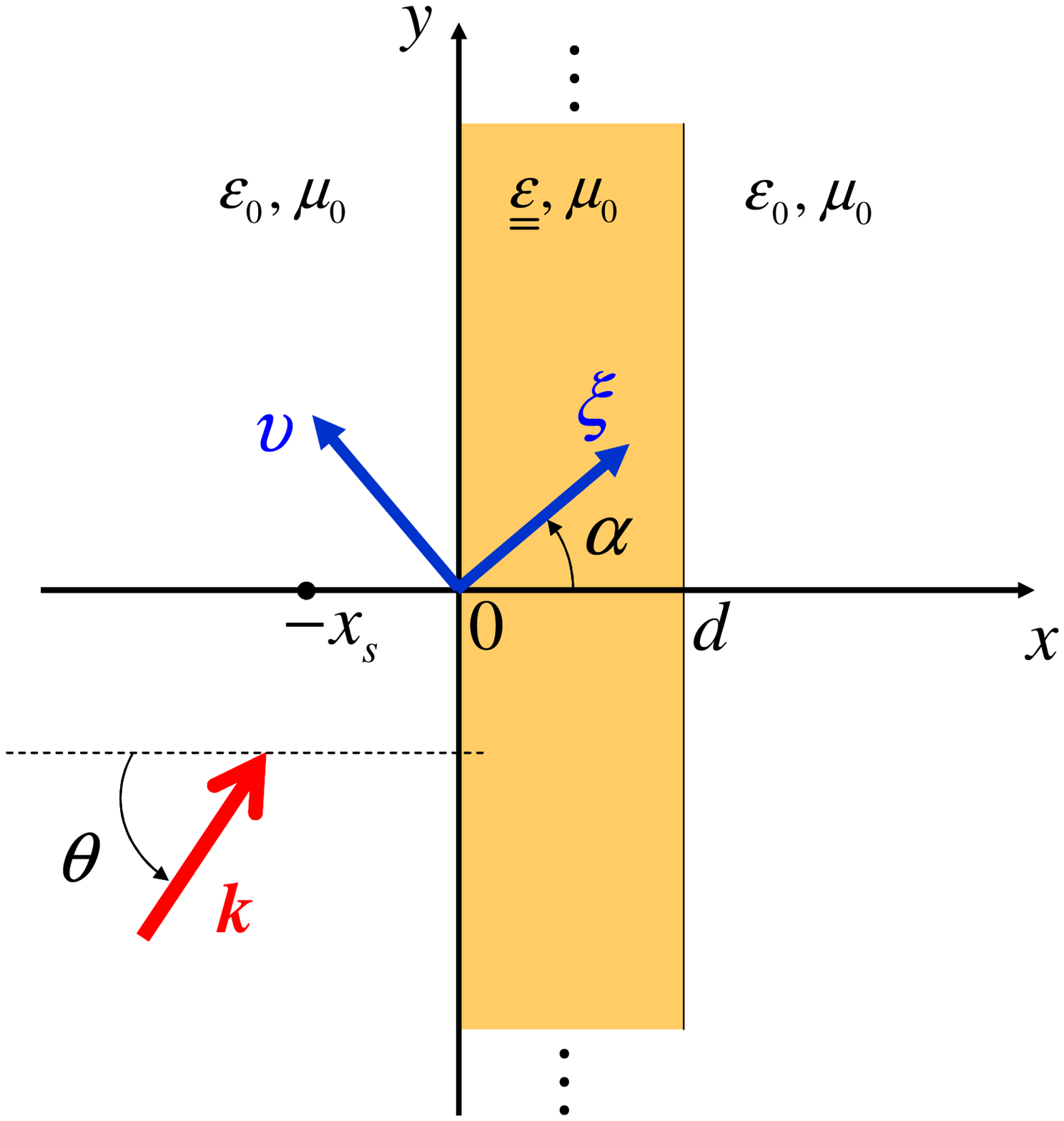}
\end{center}
\caption{(Color online) Problem schematic: a magnetic-current line source is placed at a distance $x_s$ from a uniaxially-anisotropic metamaterial slab of thickness $d$ and permittivity tensor ${\underline {\underline \varepsilon}}$ given in (\ref{eq:tuniaxial}), immersed in vacuum. Also shown are the global ($x,y$) and rotated ($\xi,\upsilon$) Cartesian coordinate reference systems, as well as the wavevector ${\bf k}$ of a propagating plane wave.}
\label{Figure1}
\end{figure}

Figure \ref{Figure1} shows the geometry of the problem. We consider a 2-D scenario featuring a metamaterial slab of thickness $d$, infinitely extent in the $y, z$ directions, immersed in vacuum. The metamaterial is assumed to be nonmagnetic and uniaxially-anisotropic with optical axis tilted of an angle $\alpha$ with respect to the $x$-axis. Accordingly, the corresponding permittivity tensor is most naturally described in the rotated (principal) reference system ($\xi,\upsilon,z$) 
\beq
{\underline {\underline \varepsilon}}_{\left\{\xi,\upsilon,z\right\}}=\varepsilon_0
\left[
\begin{array}{ccc}
\varepsilon_{\xi} & 0 & 0\\
0 & \varepsilon_{\upsilon} & 0\\
0 & 0 & \varepsilon_{\upsilon}
\end{array}
\right],
\label{eq:tuniaxial}
\eeq
and represents a rather general {\em homogenized} (effective-medium-theory) model for diverse classes of artificial materials, such as multilayered structures or wire media. In what follows, we assume that the homogenized constitutive parameters in (\ref{eq:tuniaxial}) do not depend on the wavevector. While acknowledging the implied limitations in predicting {\em nonlocal} effects that can take place in metallo-dielectric multilayers (see, e.g., Refs. \onlinecite{Elser,Belov5,Castaldi}) or wire media (see, e.g., Refs. \onlinecite{Belov8,Simovski1,Pollard}), we focus here on this simplified model which is amenable to analytical treatment. 

\begin{figure}
\begin{center}
\includegraphics [width=12cm]{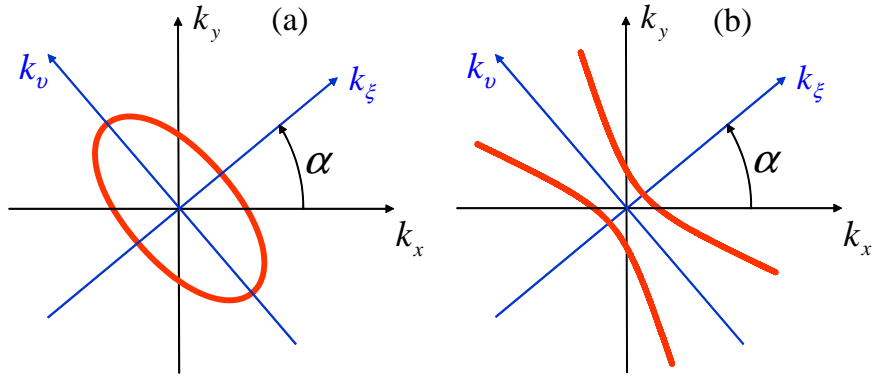}
\end{center}
\caption{(Color online) Typical elliptic (a) and hyperbolic (b) Equi-frequency contours (EFCs) pertaining to the dispersion relation in (\ref{eq:EFC}),
for $\varepsilon_{\xi}>\varepsilon_{\upsilon}>0$ and $\varepsilon_{\xi}<0, \varepsilon_{\upsilon}>0$, respectively, in the global $(k_x,k_y)$ and rotated $(k_{\xi},k_{\upsilon})$ spectral reference systems.}
\label{Figure2}
\end{figure}

The essential kinematical (wavevector, group velocity) properties of wave propagation in such a medium may be qualitatively understood by looking at the {\em equi-frequency contours} (EFCs). Once again, these are most easily studied in the spectral-wavenumber plane ($k_{\xi},k_{\upsilon}$) associated with the rotated reference coordinate systems ($\xi,\upsilon$) in Fig. \ref{Figure1}, where they assume the canonical form
\beq
\frac{k_{\xi}^2}{\varepsilon_{\upsilon}}+\frac{k_{\upsilon}^2}{\varepsilon_{\xi}}=k_0^2,
\label{eq:EFC}
\eeq
with $k_0=\omega\sqrt{\varepsilon_0\mu_0}=2\pi/\lambda_0$ denoting the vacuum wavenumber (and $\lambda_0$ the corresponding wavelength). At variance with the {\em circular} shape exhibited by isotropic media, depending on the sign of $\varepsilon_{\xi}$ and $\varepsilon_{\upsilon}$, these EFCs may be either {\em elliptic} ($\varepsilon_{\xi}\varepsilon_{\upsilon}>0$) or {\em hyperbolic} ($\varepsilon_{\xi}\varepsilon_{\upsilon}<0$), as exemplified in Fig. \ref{Figure2}. This implies, especially in the hyperbolic case [cf. Fig. \ref{Figure2}(b)], that spectral components characterized by large transverse wavevenumbers (which would be otherwise evanescent) may actually propagate. Moreover, in the ENZ limit ($\varepsilon_{\upsilon}\rightarrow 0$) of interest, the EFCs tend to become much flatter along the $k_{\upsilon}$ direction, thereby implying that the allowed group velocities (normal to the EFCs) tend to be directed along the $\xi$ direction, so that sub-wavelength details can be transported along the optical-axis direction.
Such condition (in conjunction with $\varepsilon_{\xi}\rightarrow\infty$), which was first proposed in Ref. \onlinecite{Ramakrishna1} in order to mitigate the loss-induced limitations in single-layer silver superlenses, may be attained for multilayered structures from standard mixing formulas \cite{Sihvola} using constituent materials with opposite-signed permittivities (e.g., noble metals and dielectrics, at optical frequencies). This condition was also exploited in Ref. \onlinecite{Salandrino} for far-field sub-wavelength imaging.

The reader is referred to Refs. \onlinecite{Belov2,Jin1} for different extreme-anisotropy-based sub-wavelength imaging systems relying on Fabry-Perot resonance effects.

\subsection{Green's Function}
From the above observations, we expect that the uniaxial ENZ metamaterial slab of interest may exhibit sub-wavelength image formation and lateral-displacement capabilities. In what follows, we analytically study the response to a unit-amplitude, time-harmonic [$\exp(-i\omega t)$], $z-$directed magnetic-current ($V/m^2$) line source located at a distance $x_s$ (typically $\ll\lambda_0$) from the slab (cf. Fig. \ref{Figure1})
\beq
M_z(x,y)=\delta\left(x+x_s\right)\delta\left(y\right).
\eeq
We start recalling the well-known Green's function of vacuum for the $z-$directed magnetic field,
\begin{subequations}
\begin{eqnarray}
G^{(H)}_{0}(x,y;x_s)&=&-\frac{\omega\varepsilon_0}{4}\mbox{H}_0^{(1)}\left[
k_0\sqrt{(x+x_s)^2+y^2}
\right]\\
&=&
-\frac{\omega\varepsilon_0}{4\pi}\int\limits_{-\infty}^{\infty}
\frac{\exp\left[
i\left(k_x \left|x+x_s\right|+k_y y\right)
\right]}
{k_x}
dk_y,
\label{eq:spectral}
\end{eqnarray}
\label{eq:G0}
\end{subequations}
\!\!in terms of the zeroth-order Hankel function of first kind $\mbox{H}_0^{(1)}$ (cf. Sec. 9 in \onlinecite{Abramowitz}), or the corresponding spectral-integral representation, \cite{Felsen} with
\beq
k_x=\sqrt{k_0^2-k_y^2},~~\mbox{Im}(k_x)\ge 0.
\label{eq:kx}
\eeq
Accordingly, the presence of the metamaterial slab can be accounted for by introducing in the spectral integral representation (\ref{eq:spectral}) the corresponding transverse-magnetic plane-wave transmission-coefficient $T(k_y)$ and an appropriate displacement along the $x$ direction, viz.,
\beq
G^{(H)}_S\left(x,y;x_s\right)=-
\frac{\omega\varepsilon_0}{4\pi}\int\limits_{-\infty}^{\infty}
\frac{T(k_y)}{k_x}
\exp\left[i\left(k_x \left|x+x_s-d\right|+k_y y\right)\right]dk_y.\\
\label{eq:Gs}
\eeq
The spectral integral in (\ref{eq:Gs}) cannot be generally calculated analytically in closed form. In certain regimes (see, e.g., Refs. \onlinecite{Wood,Liu1}), closed-form near-field approximations may be worked out by applying Cauchy's residue theorem and neglecting the branch-cut contribution. However, in general, a brute-force numerical integration is needed.

\subsection{ENZ Regime}
In the limit $\varepsilon_{\upsilon}\rightarrow 0$ of interest, the transmission coefficient to be considered in (\ref{eq:Gs}) reduces to (see Appendix \ref{AppA} for details)
\beq
T\left(k_y
\right)=\frac
{2\varepsilon_{\xi}k_x\exp\left(-ik_yd\tan\alpha\right)}
{2\varepsilon_{\xi}k_x-id\varepsilon_{\xi}k_0^2+idk_y^2\sec^2\alpha},
\label{eq:TENZ}
\eeq
from which it can be observed that total-transmission (i.e., perfect impedance matching) can only be achieved for
\beq
k_y=\pm k_0 \sqrt{\varepsilon_{\xi}}\cos\alpha.
\eeq
The above condition can be satisfied for propagating waves impinging from vacuum (i.e., $|k_y|\le k_0$) with incidence angle (see Fig. \ref{Figure1}) 
\beq
\theta=\pm\arcsin\left(
\sqrt{\varepsilon_{\xi}}\cos\alpha
\right),
\eeq
which admits real solutions for positive values of $\varepsilon_{\xi}$ and $\sqrt{\varepsilon_{\xi}}\left|
\cos\alpha\right|\le1$.

The transmission coefficient in (\ref{eq:TENZ}) can be recast in a convenient canonical form
\beq
T(k_y)=\frac{\Lambda k_x\exp\left(-ik_yd\tan\alpha\right)}{\left(k_x-\kappa_1\right)\left(k_x-\kappa_2\right)},
\label{eq:Tcan}
\eeq
where
\begin{subequations}
\beq
\Lambda=\frac{2i\varepsilon_{\xi}\cos^2\alpha}{d},
\eeq
\beq
\kappa_{1,2}=-\left(\frac{\cos^2\alpha}{d}\right)\left[
i\varepsilon_{\xi}\mp
\sqrt{k_0^2d^2\sec^2\alpha\left(\sec^2\alpha-\varepsilon_{\xi}\right)-\varepsilon_{\xi}^2}
\right].
\label{eq:kappaa}
\eeq
\label{eq:parENZ}
\end{subequations}
Accordingly, by substituting (\ref{eq:Tcan}) into (\ref{eq:Gs}), the slab response is reduced to calculating the canonical spectral integral
\beq
G^{(H)}_S\left(x,y;x_s\right)=
-\frac{\Lambda\omega\varepsilon_0}{4\pi}\int\limits_{-\infty}^{\infty}
\frac{\exp\left[i\left(k_xx_d+k_yy_d\right)\right]}{\left(k_x-\kappa_1\right)\left(k_x-\kappa_2\right)}dk_y,
\label{eq:Gscan}
\eeq
with (\ref{eq:kx}) and
\beq
x_d=x+x_s-d,~~y_d=y-d\tan\alpha.
\label{eq:xyd}
\eeq

\section{Analytical Results}
\label{Sec:Analytic}

\subsection{General Solution}
Introducing the (normalized) polar coordinates
\beq
\zeta=k_0\sqrt{x_d^2+y_d^2},~~\varphi=\arctan\left(\frac{y_d}{x_d}\right),
\label{eq:polar}
\eeq
it can be shown (see Appendix \ref{AppC} for details) that the canonical spectral integral in (\ref{eq:Gscan}) admits closed-form solutions of the type
\beq
G^{(H)}_S(x,y;x_s)=-\frac{\Lambda\omega\varepsilon_0}{4\pi
(\kappa_1-\kappa_2)}\left[F_1(\zeta,\varphi)-F_2(\zeta,\varphi)\right],
\label{eq:GenSol1}
\eeq
where
\begin{eqnarray}
F_{1,2}(\zeta,\varphi)&=&\exp\left(a_{1,2}^+\zeta\right)\left[
A^+_{1,2}+\chi_{1,2}^+{\cal H}\mbox{e}_0^{(1)}\left(
a_{1,2}^+,\eta_{1,2}^+,\zeta
\right)
\right]\nonumber\\
&+&
\exp\left(a_{1,2}^-\zeta\right)\left[
A^-_{1,2}+\chi_{1,2}^-{\cal H}\mbox{e}_0^{(1)}\left(
a_{1,2}^-,\eta_{1,2}^-,\zeta
\right)
\right].
\label{eq:F12}
\end{eqnarray}
In (\ref{eq:F12}), $A^+_{1,2}$ and $A^-_{1,2}$ are integration constants,  
\beq
a_{1,2}^{\pm}=
\frac{i\left(\kappa_{1,2}\cos\varphi\pm
\left|\sin\varphi\right|\sqrt{k_0^2-\kappa_{1,2}^2}
\right)}{k_0},~~\mbox{Im}\left(\sqrt{k_0^2-\kappa_{1,2}^2}\right)\ge0,
\label{eq:apar}
\eeq
\beq
\chi_{1,2}^{\pm}=
\frac{\pi \kappa_{1,2}\left(
\cos\varphi\sqrt{k_0^2-\kappa_{1,2}^2}\mp
\left|\sin\varphi\right|\kappa_{1,2}
\right)}{2ik_0\sqrt{k_0^2-\kappa_{1,2}^2}},
\label{eq:GenSol4}
\eeq
are known parameters depending on the frequency, slab parameters, and observation direction, and
\beq
{\cal H}\mbox{e}_0^{(1)}\left(
a,\eta,\zeta
\right)=\int_{\eta}^{\zeta}\exp\left(-a\tau\right)\mbox{H}_0^{(1)}(\tau)d\tau
\label{eq:CILHI}
\eeq
denotes the complementary incomplete Lipschitz-Hankel integral (CILHI) of the Hankel type\cite{Zhu} -- a special function belonging to the general class of incomplete cylindrical function. \cite{Agrest} These functions have insofar found applications in other areas of electromagnetics, ranging from diffraction to traveling-wave sources (see, e.g., Ref. \onlinecite{Dvorak0} for a review)

The lower limit of integration $\eta$ in (\ref{eq:CILHI}) is chosen so as to guarantee convergence at infinity in the complex $\tau$ plane, viz.,
\beq
\eta=\left\{
\begin{array}{ll}
\infty,~~~\mbox{Re}(a)\ge 0,\\
\infty\exp(i\pi),~~~\mbox{Re}(a)< 0.
\end{array}
\right.
\eeq

The calculation of the $\varphi$-dependent integration constant $A^+_{1,2}$ and $A^-_{1,2}$ in (\ref{eq:F12}) is generally quite involved (see the discussion in Appendix \ref{AppD}). Nevertheless, for the near-field configuration of direct interest to this investigation, featuring a source located very close to the input slab interface ($x_s\ll\lambda_0$), this calculation becomes remarkably simpler for two specific observation planes located in the two principal planes of the image, namely: transverse cuts at the output slab interface ($x=d$), and orthogonal cuts passing through the fiducial image position ($y_{\alpha}=d\tan\alpha$). For the former case ($x=d$), we obtain (see Appendix \ref{AppD} for details) 
\begin{subequations}
\begin{eqnarray}
A^-_{1,2}&=&0,\label{eq:Am}
\\
A^+_{1,2}&=&C_{1,2}
-\chi_{1,2}^+{\cal H}\mbox{e}_0^{(1)}\left(a_{1,2}^+,\eta_{1,2}^+,0\right)
-\chi_{1,2}^-{\cal H}\mbox{e}_0^{(1)}\left(a_{1,2}^-,\eta_{1,2}^-,0\right),
\label{eq:Ap}
\end{eqnarray}
\label{eq:IC1}
\end{subequations}
where
\begin{eqnarray}
C_{1,2}&=&-\frac{i\kappa_{1,2}}{\sqrt{k_0^2-\kappa_{1,2}^2}}
\left\{
2\mbox{arctan}\left(\displaystyle{\frac{\kappa_{1,2}}{\sqrt{k_0^2-\kappa_{1,2}^2}}}\right)\right.\nonumber\\
&+&
\Biggl.
\pi\left\{
1-4 u\left[\mbox{Re}\left(\kappa_{1,2}\right)\right]u\left[\mbox{Im}\left(\kappa_{1,2}\right)\right]
\right\}
\Biggr\},
\label{eq:C12}
\end{eqnarray}
with $u$ denoting the Heaviside unit-step function, and\cite{Zhu}
\beq
{\cal H}\mbox{e}_0^{(1)}\left(a,\eta,0\right)=\frac{2i\log\left(a+\sqrt{a^2+1}\right)-\pi}{\pi\sqrt{a^2+1}},
\label{eq:CILHI0}
\eeq
with the principal branch chosen for the natural logarithm, and the branch-cut for $\sqrt{a^2+1}$ chosen as
\beq
\left\{
\begin{array}{ll}
\mbox{Re}\left(\sqrt{a^2+1}\right)\ge 0,~~\forall a,\\
\mbox{Im}\left(\sqrt{a^2+1}\right)\ge 0,~~\mbox{for}~\mbox{Re}\left(\sqrt{a^2+1}\right)= 0.
\end{array}
\right.
\eeq
For the latter case ($y=d\tan\alpha$), the above expressions are still valid if $\mbox{Re}(a_{1,2}^+)=\mbox{Re}(a_{1,2}^-)<0$, otherwise they simply reduce to (see Appendix \ref{AppD} for details)
\beq
A_{1,2}^+=A_{1,2}^-=0.
\label{eq:A120}
\eeq

It is rather remarkable that in these two relevant planes, of most interest to analyze the imaging properties of the metamaterial slabs, we are able to obtain a closed-form analytical solution of the field distribution induced by an arbitrary magnetic source. This result will be used in the following to highlight the remarkable imaging properties of ENZ metamaterials with arbitrarily tilted anisotropy axis.

\subsection{Physical Interpretation}

It is of particular interest to identify the distinct phenomena 
involved in the wave interaction with the geometry of Fig. \ref{Figure1} as a function of the various parameters considered
in our analytical solution in (\ref{eq:GenSol1}) and (\ref{eq:F12}). First, we notice that the exponential terms $\exp\left(a_{1,2}^{\pm}\zeta\right)$ account for the dominant resonances exhibited by the slab, corresponding to pole singularities in the spectral integral formulation. \cite{Felsen}
Their localization properties can be readily related to the slab physical parameters via (\ref{eq:apar}) and (\ref{eq:kappaa}).

The terms $\exp\left(a_{1,2}^{\pm}\zeta\right){\cal H}\mbox{e}_0^{(1)}\left(a_{1,2}^{\pm},\eta_{1,2}^{\pm},\zeta\right)$ instead account for the continuous radiation spectrum, corresponding to the branch-cut contributions in the spectral integral formulation. \cite{Felsen} We remark that, unlike the configuration in Refs. \onlinecite{Wood,Liu1}, these contributions are generally non-negligible. Recalling the definition in (\ref{eq:CILHI}), these terms may be physically interpreted as smoothed versions (via convolution with a complex-exponential window) of a zeroth-order Hankel function of the first kind $\mbox{H}_0^{(1)}$. 

\subsection{Computational Aspects}

\subsubsection{Numerical Computation of CILHIs}
\label{Sec:NumCILHI}
Recalling the expression of the $a_{1,2}^{\pm}$ parameters in (\ref{eq:apar}), it is evident that the numerical implementation of our analytical solution requires in general the calculation of complex-argument CILHIs of the Hankel type. In Ref. \onlinecite{Zhu},
various series expansions were derived for accurate and efficient computation of these special functions. In particular, by comparison against brute-force numerical integration (via adaptive Gauss/Kronrod quadrature) of the corresponding spectral integrals, efficiency improvements ranging from one to nearly three orders of magnitudes were found.

Our numerical implementation is based on a selective application of the factorial-Neumann and Struve-function series expansions, following the guidelines of Ref. \onlinecite{Zhu} for the various parameter ranges.
Although a proper numerical optimization to maximize the calculation efficiency of these coefficients goes beyond our interest, it is evident that our closed-form analytical solution may significantly outperform conventional numerical solvers.

\subsubsection{Small-Argument Approximation}
Since the application of the metamaterial slab and the focus of this paper is concentrated on the
near-field sub-wavelength imaging scenario, with $x_s\ll\lambda_0$ and $x=d$, a simple small-argument ($\zeta\ll 1$) analytical approximation for the CILHIs may be conveniently utilized. First, we recall the small-argument approximation of the zeroth-order Hankel function of the first kind  (cf. Eqs. (9.1.12) and (9.1.13) in Ref. \onlinecite{Abramowitz}),
\beq
\mbox{H}_0^{(1)}\left(\tau\right)\sim 1+\left(\frac{2i}{\pi}\right)
\left[
\log\left(\frac{\tau}{2}\right)+\gamma
\right],
\eeq
where $\gamma$ denotes the Euler-Mascheroni constant. \cite{Abramowitz} Using this approximation, it follows that
\begin{eqnarray}
\exp\left(a\zeta\right)\int_{0}^{\zeta} \exp\left(-a\tau\right)\mbox{H}_0^{(1)}\left(\tau\right)d\tau&\sim&
\left[\frac{\exp\left(a\zeta\right)-1}{a}\right]-\frac{2i}{\pi a}\log\left(\frac{\zeta}{2}\right)\nonumber\\
&-&\frac{2i\left\{\exp\left(a\zeta\right)
\left[
E_1\left(a\zeta\right)+\log\left(2a\right)
\right]+\gamma
\right\}}{\pi a},
\end{eqnarray}
where $E_1$ denotes the exponential integral [cf. Eq. (5.1.1) in Ref. \onlinecite{Abramowitz}] which, recalling the expansion in Eq. (5.1.11) of Ref. \onlinecite{Abramowitz} and exploiting a (1,1) Pad\'e approximant \cite{Baker} for the power-series part, 
may be conveniently approximated as
\beq
E_1\left(a\zeta\right)\sim \frac{4a\zeta}{4+a\zeta}-\gamma-\log\left(a\zeta\right).
\eeq
Recalling (\ref{eq:CILHI}) and (\ref{eq:CILHI0}), we finally obtain the small-argument approximation
\begin{eqnarray}
\exp(a\zeta){\cal H}\mbox{e}_0^{(1)}\left(a,\eta,\zeta\right)&\sim&
\exp(a\zeta)\left[
\frac{2i\log\left(a+\sqrt{a^2+1}\right)-\pi}{\pi\sqrt{a^2+1}}
-\frac{8ia\zeta}{\pi a\left(4+a\zeta\right)}
\right]\nonumber\\
&+&\frac{\left[\exp\left(a\zeta\right)-1\right]
\left\{2i
\left[\log\left(\displaystyle{\frac{\zeta}{2}}\right)+\gamma\right]+\pi
\right\}
}{\pi a},
\label{eq:NFA}
\end{eqnarray}
in terms of simple analytical functions. By substituting (\ref{eq:NFA}) in (\ref{eq:F12}), we obtain
\begin{eqnarray}
F_{1,2}(\zeta,\varphi)&\sim&A_{1,2}^+\exp\left(a_{1,2}^+\zeta\right)+\chi_{1,2}^+\frac{\left[\exp\left(a_{1,2}^+\zeta\right)-1\right]
\left\{2i
\left[\log\left(\displaystyle{\frac{\zeta}{2}}\right)+\gamma\right]+\pi
\right\}
}{\pi a_{1,2}^+}\nonumber\\
&+&\chi_{1,2}^+\exp(a_{1,2}^+\zeta)\left[
\frac{2i\log\left(a_{1,2}^++\sqrt{(a_{1,2}^+)^2+1}\right)-\pi}{\pi\sqrt{(a_{1,2}^+)^2+1}}
-\frac{8ia_{1,2}^+\zeta}{\pi a_{1,2}^+\left(4+a_{1,2}^+\zeta\right)}
\right]\nonumber\\
&+&\chi_{1,2}^-\frac{\left[\exp\left(a_{1,2}^-\zeta\right)-1\right]
\left\{2i
\left[\log\left(\displaystyle{\frac{\zeta}{2}}\right)+\gamma\right]+\pi
\right\}
}{\pi a_{1,2}^-}\nonumber\\
&+&\chi_{1,2}^-\exp(a_{1,2}^-\zeta)\left[
\frac{2i\log\left(a_{1,2}^-+\sqrt{(a_{1,2}^-)^2+1}\right)-\pi}{\pi\sqrt{(a_{1,2}^-)^2+1}}
-\frac{8ia_{1,2}^-\zeta}{\pi a_{1,2}^-\left(4+a_{1,2}^-\zeta\right)}
\right],
\label{eq:F12app}
\end{eqnarray}
which, substituted in (\ref{eq:GenSol1}), yield the final approximation for the Green's function (not given explicitly here for brevity).  Once implemented, this solution may provide a complete description of the imaging properties of the metamaterial slab, based on conventional basic functions. Its overall applicability and accuracy, which is expected to be restricted to small values of the argument, will be quantitatively assessed in Sec. \ref{Sec:NumRes}.

\section{Representative Numerical Results and Physical Insights}
\label{Sec:Results}

\subsection{Generalities}
\label{Sec:ResultsGen}

In what follows, for certain representative ENZ parameter configurations, we validate and calibrate the analytical solutions derived in Sec. \ref{Sec:Analytic} against a reference solution obtained via brute-force numerical quadrature of the spectral integral in (\ref{eq:Gs}) with the general transmission coefficient in (\ref{eq:TT}). We further discuss the imaging properties of the metamaterial slab for specific design parameters of interest.
Our reference solution relies on Gaussian-type quadrature rules (cf. Sec. 25.4 in Ref. \onlinecite{Abramowitz}), with the number of nodes refined adaptively so as to guarantee a four-digit accuracy. It is worth pointing out that no particular attempt was made to optimize its numerical implementation, since a careful and thorough assessment of the computational convenience of CILHI-based solutions vs. numerical spectral integration was already carried out in Ref. \onlinecite{Zhu} (see also the discussion in Sec. \ref{Sec:NumCILHI}), and is therefore not the focus of our investigation here.

\subsection{Numerical Results and Discussion}
\label{Sec:NumRes}
\begin{figure}
\begin{center}
\includegraphics [width=16cm]{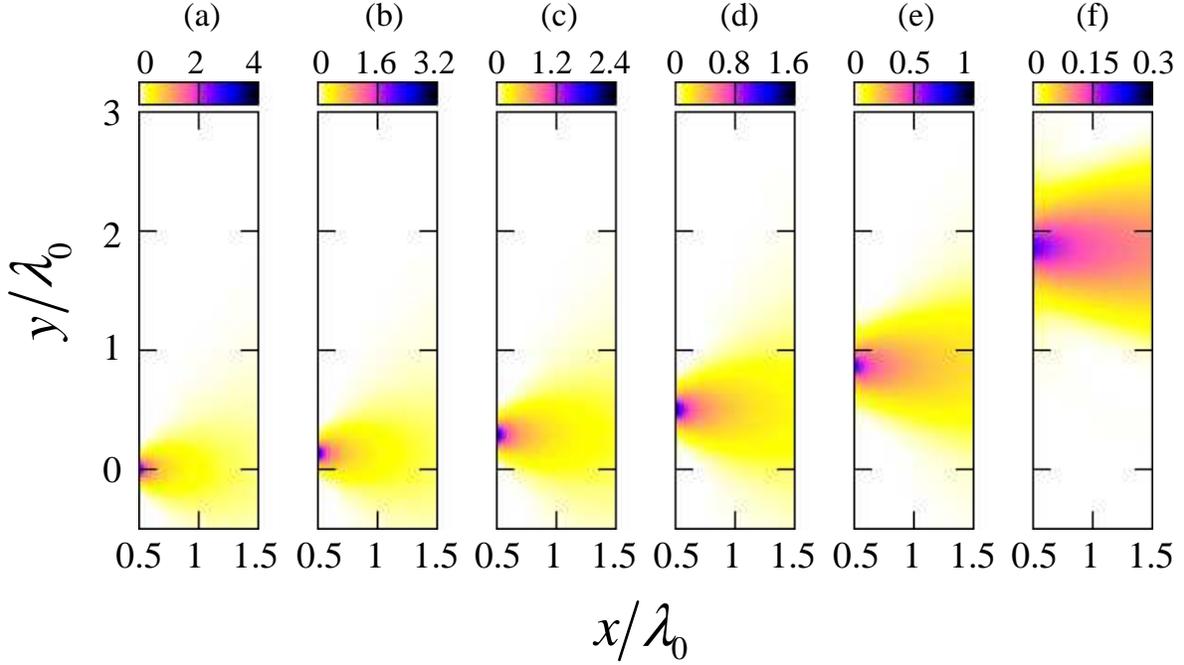}
\end{center}
\caption{(Color online) Normalized Green's function intensity maps (within and beyond the image plane $x=d$), for a source located $x_s=\lambda_0/50$ away from a slab of thickness $d=0.5\lambda_0$, $\varepsilon_{\upsilon}=0$, $\varepsilon_{\xi}=-2$, and the optical-axis angle $\alpha$ varying from 0 (a) to $75^o$ (f) with step of $15^o$, computed via the reference solution [cf. \ref{eq:Gs}]. Here and henceforth, intensities are normalized with respect to the peak-intensity at the image plane in the absence of the slab, i.e., $\left|G^{(H)}_0(d,0,x_s)\right|^2$ [cf. (\ref{eq:G0})].}
\label{Figure3}
\end{figure}

We start considering an ideal lossless configuration featuring a line source placed at a distance $x_s=\lambda_0/50$ away from a slab of thickness $d=0.5\lambda_0$, and constitutive parameters $\varepsilon_{\upsilon}=0$, $\varepsilon_{\xi}=-2$. Figure \ref{Figure3} shows the normalized Green's function intensity maps, within and beyond the image plane $x=d$, computed by using the reference solution [cf. (\ref{eq:Gs})], for values of the optical-axis angle $\alpha$ ranging from $0$ to $75^o$.
Similarly to what observed in Ref. \onlinecite{Wang}, the images are maximally localized at the image plane $x=d$ and are subject to a lateral displacement (of a quantity $y_{\alpha}=d\tan\alpha$) with respect to the source position ($y=0$), which is in agreement with the fact that sub-wavelength details of the source are effectively transported with small distortion along the optical-axis direction. As it can be observed, increasing values of the tilt angle $\alpha$ (i.e., of the lateral displacement) are accompanied by progressively worse localization properties (note the different color scales of the plots).

\begin{figure*}
\begin{center}
\includegraphics [width=15cm]{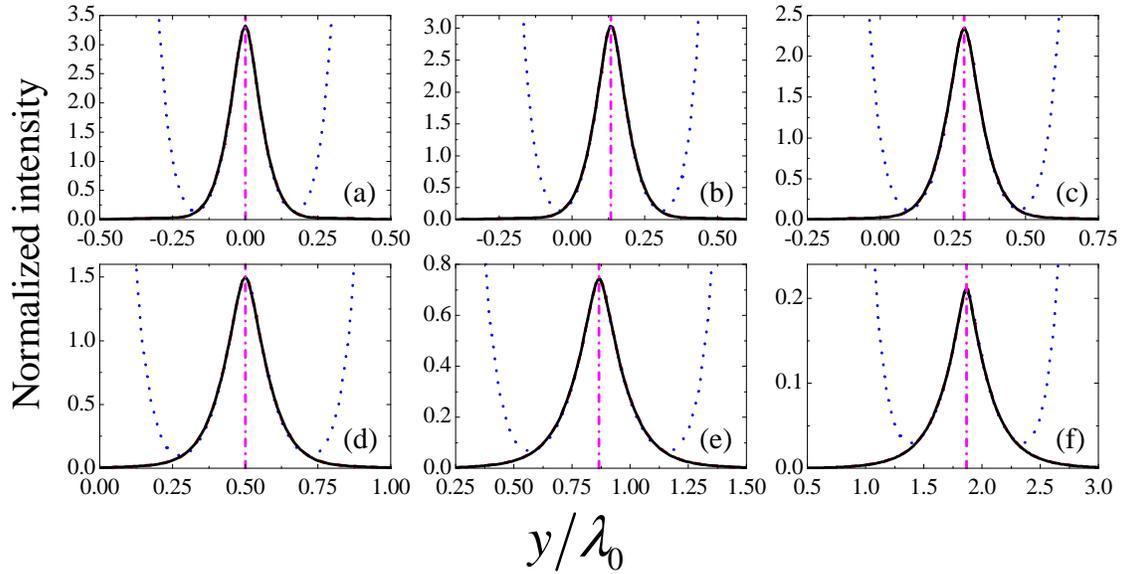}
\end{center}
\caption{(Color online) As in Fig. \ref{Figure3}, but lateral cuts at the image plane $x=d$ (black-solid curves) compared with the ``exact'' and small-argument-approximated CILHI-based analytical solutions (red-dashed and blue-dotted curves, respectively). The magenta-dashed-dotted vertical lines indicate the fiducial positions $y_{\alpha}=d\tan\alpha$ of the laterally-displaced images.}
\label{Figure4}
\end{figure*}

\begin{figure*}
\begin{center}
\includegraphics [width=15cm]{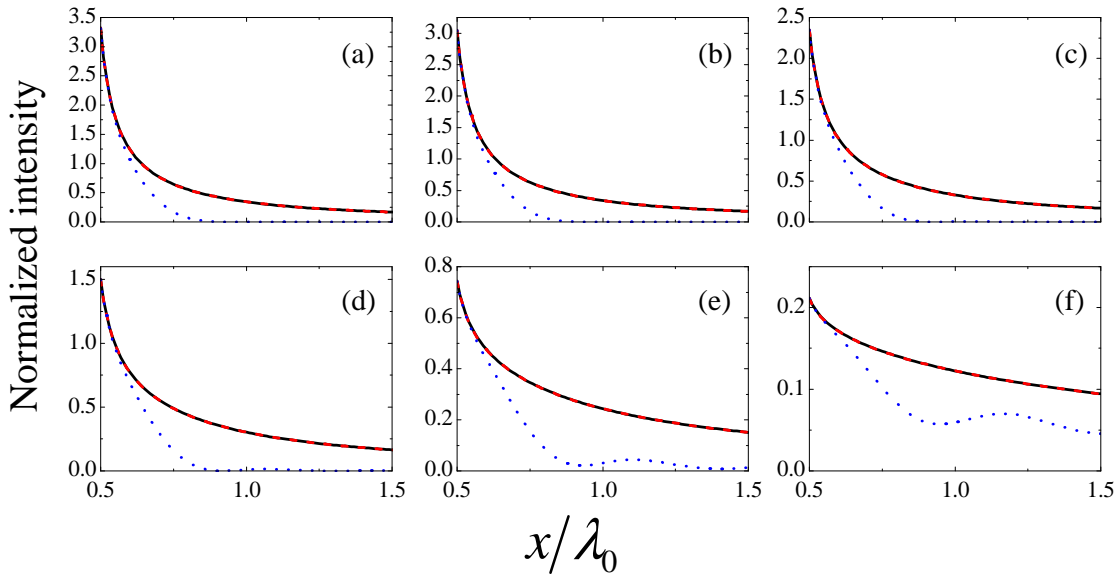}
\end{center}
\caption{(Color online) As in Fig. \ref{Figure4}, but orthogonal cuts at the fiducial position of the image $y_{\alpha}=d\tan\alpha$.}
\label{Figure5}
\end{figure*}

Figures \ref{Figure4} and \ref{Figure5} show the intensity profiles along the two principal cuts, namely, the image plane at the slab interface ($x=d$) and the orthogonal plane passing trough the fiducial image position ($y_{\alpha}=d\tan\alpha$), respectively, comparing the reference solution [cf. (\ref{eq:Gs}), black-solid curves] with the CILHI-based analytical solutions in (\ref{eq:GenSol1}) and (\ref{eq:F12}) [with the parameters $\Lambda$ and $\kappa_{1,2}$ given by (\ref{eq:parENZ})]. 
Along these cuts, it was possible to apply the small-argument approximation and derive Eqs. (\ref{eq:NFA}).
More specifically, both the ``exact'' (i.e., numerically computed as in Sec. \ref{Sec:NumCILHI}) and small-argument-approximated [cf. (\ref{eq:NFA})] CILHIs are considered (red-dashed and blue-dotted curves, respectively). The ``exact'' CILHI-based solutions are practically indistinguishable from the reference solutions [cf. (\ref{eq:Gs})] on the scale of the plots, whereas, as expected, the small-argument approximation breaks down away from the peak. These limitations are more evident in the orthogonal cuts (Fig. \ref{Figure5}), since the longitudinal localization is typically poorer.  It is important to stress that, compared to a conventional focusing lens, the metamaterial slab is able to focus the transverse image well below the diffraction limit, but it is not very effective in focusing in the longitudinal plane. Effectively, as shown in Fig. \ref{Figure5}, the sub-wavelength spot decays away from the slab with a conventional exponential drop, due to the diffraction of the high-wavenumber spectral components of the image transported to the back of the slab. 

\begin{figure}
\begin{center}
\includegraphics [width=10cm]{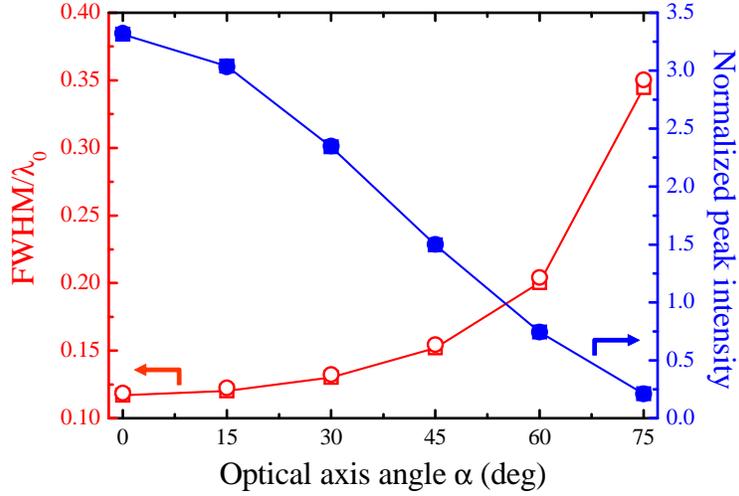}
\end{center}
\caption{(Color online) Parameters as in Fig. \ref{Figure4}. FWHM (empty markers, left axis) and peak-intensity (full markers, right axis) at the image plane $x=d$, estimated via the reference solution [cf. (\ref{eq:Gs}), circular markers] and the small-argument-approximated CILHI-based analytical solution (square markers), as a function of the optical-axis angle $\alpha$.}
\label{Figure6}
\end{figure}

It is interesting to notice in Figs. \ref{Figure4} and \ref{Figure5} how
the small-argument approximation is sufficient to correctly capture the essential localization properties of the image, especially in the transverse direction.
In what follows, for quantitative assessments, we consider two typical figures of merit: the {\em full-width-at-half-maximum} (FWHM) and the (normalized) peak intensity at the image plane $x=d$.
Figure \ref{Figure6} compares the FWHM and (normalized) peak-intensity, estimated via the reference solution [cf. (\ref{eq:Gs}), circular markers] and the small-argument CILHI-based analytical solution (square markers), as a function of $\alpha$. 
Overall, a uniformly good agreement is observed, with maximum errors $<2\%$ in the FWHM and $<0.3\%$ in the peak-intensity. 
Consistently with the visual impression from Figs. \ref{Figure3} and \ref{Figure4}, both observables deteriorate with increasing values of $\alpha$. More specifically, over the range $0\le\alpha\le 75^o$, the FWHM increases from $\sim 0.12 \lambda_0$ to $\sim 0.35\lambda_0$, while the (normalized) peak-intensity decreases from $\sim 3.3$ to $\sim 0.2$. To sum up, the ENZ configuration analyzed in this paper is able to transport, and laterally displace, sub-wavelength details of a source from the input to the output slab interface, with resolutions as good as a tenth of a wavelength. However, large lateral displacements result in a larger degradation of the image resolution and intensity. For instance, lateral displacements of nearly two wavelengths [cf. Figs. \ref{Figure3}(f), \ref{Figure4}(f), and \ref{Figure5}(f)] may be attained at the expense of a factor $\sim3$ in the resolution and over an order of magnitude in the intensity.

\begin{figure}
\begin{center}
\includegraphics [width=10cm]{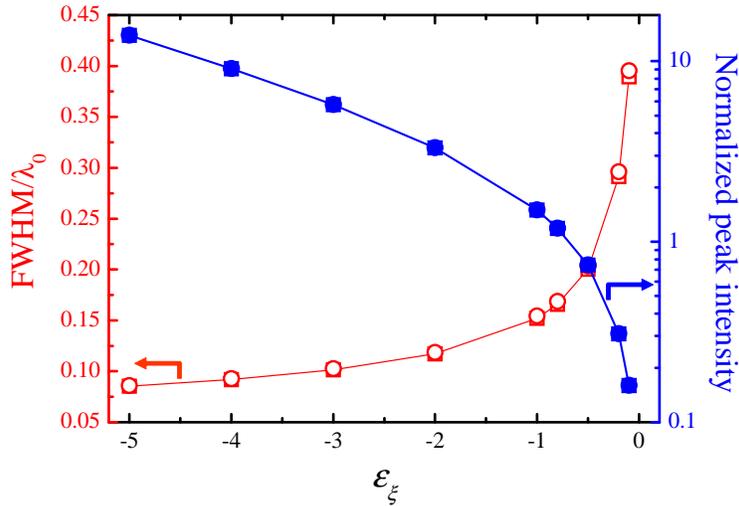}
\end{center}
\caption{(Color online) As in Fig. \ref{Figure6}, but as a function of $\varepsilon_{\xi}$, for $\alpha=0$. Note the log-scale on the right axis.}
\label{Figure7}
\end{figure}

For the same configuration, and a fixed optical-axis direction ($\alpha=0$), Fig. \ref{Figure7} shows the two considered figures of merit as a function of $\varepsilon_{\xi}$ in the hyperbolic-medium regime. As expected, recalling our discussion at the end of Sec. \ref{Sec:Gen}, the figures of merit improve for larger absolute values of $\varepsilon_{\xi}$. Conversely, they strongly deteriorate for $\varepsilon_{\xi}\rightarrow 0$. This is also not surprising, as it is well known that in the {\em isotropic} ENZ limit the slab becomes a highly-selective spatial filter. \cite{Alu0}  Also in these cases, the agreement between the reference solution [cf. (\ref{eq:Gs})] and the small-argument CILHI-based analytical solution is very good, with same maximum errors as above. 

\begin{figure}
\begin{center}
\includegraphics [width=10cm]{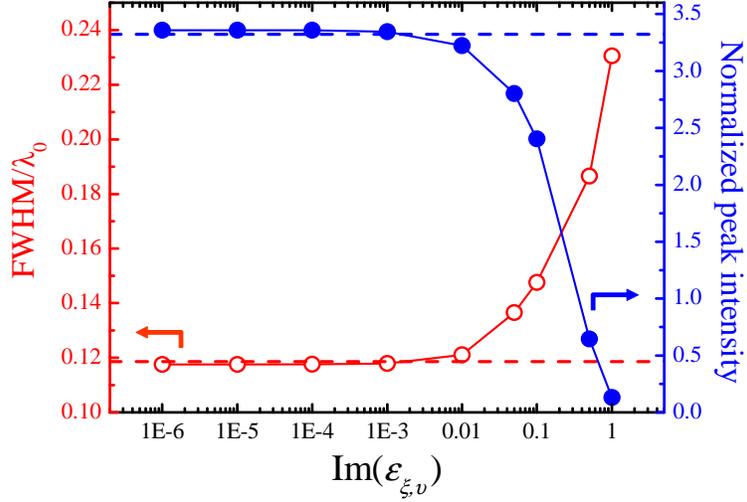}
\end{center}
\caption{(Color online) FWHM (empty markers, left axis) and peak-intensity (full markers, right axis) at the image plane $x=d$, estimated via the reference solution [cf. (\ref{eq:Gs})], for $x_s=\lambda_0/50$, $d_=0.5\lambda_0$, $\alpha=0$, $\mbox{Re}(\varepsilon_{\xi})=-2$, $\mbox{Re}(\varepsilon_{\upsilon})=10^{-3}$, as a function of $\mbox{Im}(\varepsilon_{\xi,\upsilon})$ (in log-scale). Also shown (as horizontal dashed lines) are the corresponding estimates from the small-argument-approximated CILHI-based analytical solution.}
\label{Figure8}
\end{figure}

\begin{figure}
\begin{center}
\includegraphics [width=10cm]{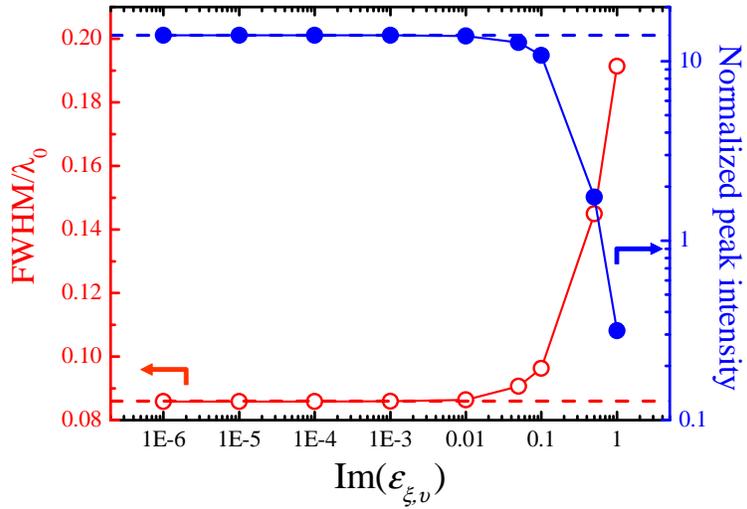}
\end{center}
\caption{(Color online) As in Fig. \ref{Figure8}, but for $\mbox{Re}(\varepsilon_{\xi})=-5$. Note also the log-scale on the right axis.}
\label{Figure9}
\end{figure}

Next, we assess the critical ENZ assumption underlying our analytical solution.
It is well known that in practical implementations (e.g., metallo-dielectric multilayers) the condition $\varepsilon_{\upsilon}=0$ may only be approximatively attained due to the presence of losses, and in any case limited to one single frequency point.
While it is possible, in principle, to achieve a vanishing real part (at a given frequency), zeroing the imaginary part is prevented by unavoidable material losses. Nevertheless, recent studies\cite{Ciattoni} have demonstrated the promising potentials of {\em gain-assisted} nanocomposites for the synthesis of artificial materials with very small values ($\sim 10^{-2}$) of the real and imaginary part of $\varepsilon_{\upsilon}$. In order to assess the applicability of our proposed analytical solution to such regime, we consider a more realistic configuration with, $d=0.5\lambda_0$, $\alpha=0$, $\mbox{Re}(\varepsilon_{\xi})=-2$, and a small but nonzero $|\varepsilon_{\upsilon}|$. More specifically, we assume $\mbox{Re}(\varepsilon_{\upsilon})=10^{-3}$, and let the imaginary parts of $\varepsilon_{\xi}$ and $\varepsilon_{\upsilon}$ vary over several decades. Figure \ref{Figure8} shows the corresponding FWHM and peak-intensity estimated via the reference solution [cf. (\ref{eq:Gs})]. As expected, for asymptotically vanishing losses, they approach the estimates from our small-argument CILHI-based analytical solution (shown as horizontal dashed lines), and progressively depart from them for increased loss levels. In particular, for $\mbox{Im}(\varepsilon_{\xi,\upsilon})=10^{-2}$, the agreement is still satisfactory, with only a $\sim 2\%$ error in the FWHM and a $\sim 3\%$ error in the peak-intensity. For $\mbox{Im}(\varepsilon_{\xi,\upsilon})=0.05$, the errors increase to $\sim 15\%$ and $\sim 16\%$, respectively, which may still be acceptable.
Qualitatively similar trends where observed for different values of $\mbox{Re}(\varepsilon_{\xi})$, as exemplified in Fig. \ref{Figure9}. In this case, pertaining to $\mbox{Re}(\varepsilon_{\xi})=-5$, the errors in the FWHM and peak-intensity are $\sim 0.4\%$ and $\sim 1\%$, respectively, for $\mbox{Im}(\varepsilon_{\xi,\upsilon})=10^{-2}$, and $\sim 5\%$ and $\sim 9\%$, respectively, for $\mbox{Im}(\varepsilon_{\xi,\upsilon})=0.05$. From the physical point of view, our results indicate that the imaging properties of the ENZ anisotropic slabs considered here are quite robust to losses and frequency variations.

Overall, the above results also indicate that our proposed analytical solution accurately captures the essential image-formation properties in the scenario of interest, and it can be safely applied in low-loss scenarios.

\section{Conclusions}
\label{Sec:Conclusions}

In this paper, we have presented an analytical study of the sub-wavelength imaging properties of uniaxially-anisotropic ENZ metamaterial slabs with tilted optical axis. In particular, we have derived a closed-form analytical solution for the 2-D
Green's function in terms of special cylindrical functions. These functions can be efficiently computed via well-established numerical schemes, yielding computational savings up to nearly three orders of magnitudes by comparison with brute-force numerical quadrature of the corresponding spectral integrals. Moreover, in the near-field parameter range of interest, they can be conveniently approximated in terms of simple analytical functions. 

Validation and calibration of our results against a numerical-integration-based reference solution [cf. (\ref{eq:Gs})] confirmed the applicability of our solution to sub-wavelength imaging scenarios with low-loss constitutive parameters that are within reach of current (e.g., gain-assisted) technologies.  We have employed this solution to analyze the imaging properties of anisotropic ENZ metamaterial slabs varying the design parameters and tilt angle.

Current and future investigations are aimed at exploiting our proposed parameterization in design procedures and optimization schemes to improve the imaging capabilities of these devices. Also of interest are possible extensions to account for spatial-dispersion (nonlocal) effects, as well as the to predict the emission enhancement for quantum sources radiating in ENZ media.\cite{Narimanov}

\appendix

\section{Pertaining to (\ref{eq:TENZ})}
\label{AppA}
The plane-wave transmission coefficient of the uniaxial slab described by the permittivity tensor in (\ref{eq:tuniaxial}) is computed by expanding the field inside the slab in terms of forward and backward plane waves (with conserved transverse wavevectors), and enforcing the phase-matching and transverse-field-continuity conditions at the interfaces. For the assumed transverse-magnetic polarization, via cumbersome but straightforward algebra, we obtain the general expression
\beq
T\left(k_y
\right)
=\frac{2\exp\left(i k_{x2}d\right)\varepsilon_{xx}k_x\left(\varepsilon_{xy}^2-\varepsilon_{xx}\varepsilon_{yy}\right)
\left(k_{x1}-k_{x2}\right)}
{U_1^-(k_y)U_2^+(k_y)-\exp\left[i\left(k_{x2}-k_{x1}\right)d\right]U_1^+(k_y)U_2^-(k_y)}
\label{eq:TT},
\eeq
where
\begin{subequations}
\begin{eqnarray}
\varepsilon_{xx}&=&\varepsilon_{\xi}\cos^2\alpha+\varepsilon_{\upsilon}\sin^2\alpha,\\
\varepsilon_{yy}&=&\varepsilon_{\xi}\sin^2\alpha+\varepsilon_{\upsilon}\cos^2\alpha,\\
\varepsilon_{xy}&=&\left(\varepsilon_{\xi}-\varepsilon_{\upsilon}\right)\sin\alpha\cos\alpha
\end{eqnarray}
\label{eq:cartesian}
\end{subequations}
represent the Cartesian components of the relative permittivity tensor in (\ref{eq:tuniaxial}), and
\beq
k_{x1,2}=\frac{-\varepsilon_{xy}k_y\mp\sqrt{\left(\varepsilon_{xx}\varepsilon_{yy}-\varepsilon_{xy}^2\right)
\left(
\varepsilon_{xx}k_0^2-k_y^2
\right)
}}{\varepsilon_{xx}},
\label{eq:kx12}
\eeq
\beq
U_{\varsigma}^{\pm}(k_y)=\varepsilon_{xy}^2k_x-\varepsilon_{xx}\left(
\varepsilon_{yy}k_x\pm k_{x\varsigma}
\right)\mp\varepsilon_{xy} k_y,~~\varsigma=1,2.
\label{eq:Upm}
\eeq

It can readily be verified that the limit $\varepsilon_{\upsilon}\rightarrow 0$ leads to a $0/0$ indeterminate form in (\ref{eq:TT}). 
In order to evaluate this limit, noting from (\ref{eq:kx12}) that $k_{x1}\rightarrow k_{x2}$, it is expedient to linearize the exponential function in the denominator of (\ref{eq:TT}), viz.,
\beq
T\left(k_y
\right)
\approx\frac{2\exp\left(i k_{x2}d\right)\varepsilon_{xx}k_x\left(\varepsilon_{xy}^2-\varepsilon_{xx}\varepsilon_{yy}\right)
\left(k_{x1}-k_{x2}\right)}
{U_1^-(k_y)U_2^+(k_y)-\left[1+i\left(k_{x2}-k_{x1}\right)d\right]U_1^+(k_y)U_2^-(k_y)}
\label{eq:TTa}.
\eeq
Next, via straightforward McLaurin expansions, we obtain for the various terms in (\ref{eq:TTa})
\beq
\varepsilon_{xx} \left(
\varepsilon_{xy}^2-\varepsilon_{xx}\varepsilon_{yy}
\right)\left(
k_{x1}-k_{x2}
\right)\sim
2\left(\varepsilon_{\xi}\varepsilon_{\upsilon}\right)^{\frac{3}{2}} \sqrt{k_0^2\varepsilon_{\xi}\cos^2\alpha-k_y^2}+
O\left(\varepsilon_{\upsilon}^{\frac{5}{2}}\right),
\label{eq:Mc1}
\eeq
\beq
U_1^-(k_y)U_2^+(k_y)-U_1^+(k_y)U_2^-(k_y)\sim
4\left(\varepsilon_{\xi}\varepsilon_{\upsilon}\right)^{\frac{3}{2}} k_x \sqrt{k_0^2\varepsilon_{\xi}\cos^2\alpha-k_y^2}+
O\left(\varepsilon_{\upsilon}^{\frac{5}{2}}\right),
\eeq
\beq
\left(k_{x2}-k_{x1}\right)U_1^+(k_y)U_2^-(k_y)\sim
2\sqrt{\varepsilon_{\xi}}
\varepsilon_{\upsilon}^{\frac{3}{2}}
\left(k_0^2\varepsilon_{\xi}\cos^2\alpha-k_y^2
\right)^{\frac{3}{2}}\sec^2\alpha+
O\left(\varepsilon_{\upsilon}^2\right),
\label{eq:Mc3}
\eeq
with $O(\cdot)$ denoting the Landau symbol. The limit in (\ref{eq:TENZ}) readily follows by substituting (\ref{eq:Mc1})--(\ref{eq:Mc3}) into (\ref{eq:TTa}), neglecting higher-order terms, and simplifying the dominant term $\varepsilon_{\upsilon}^{\frac{3}{2}}$.

\section{Derivation of the General Solution in (\ref{eq:GenSol1})--(\ref{eq:GenSol4})}
\label{AppC}
First, by rewriting the rational part of the integrand in (\ref{eq:Gscan}) as
\beq
\frac{1}{(k_x-\kappa_1)(k_x-\kappa_2)}=\frac{\kappa_1}{k_x(k_x-\kappa_1)(\kappa_1-\kappa_2)}-\frac{\kappa_2}{k_x(k_x-\kappa_2)(\kappa_1-\kappa_2)},
\label{eq:recast}
\eeq
we obtain [cf. (\ref{eq:GenSol1})] a different, generic canonical integral of the form
\beq
F(\zeta,\varphi)=\kappa\int\limits_{-\infty }^{\infty}
\frac{\exp\left[i\zeta\left(\displaystyle{\frac{k_x}{k_0}}\cos\varphi+\displaystyle{\frac{k_y}{k_0}}\sin\varphi\right)\right]}{k_x(k_x-\kappa)} dk_y
\label{eq:Fint}
\eeq
in the polar coordinates given by (\ref{eq:polar}).
In Ref. \onlinecite{Dvorak2}, closed-form calculation of spectral integrals of this type was carried out via rather cumbersome contour-integration techniques. In what follows, we rely on an alternative, relatively simpler, differential-equation-based procedure proposed in Ref. \onlinecite{Dvorak3} (based on the work in Ref. \onlinecite{Wu}).
The basic idea is to construct an inhomogeneous differential equation (in the $\zeta$ variable) satisfied by the integral in (\ref{eq:Fint}). To this aim, we apply to (\ref{eq:Fint}) a second-order differential operator
\begin{subequations}
\begin{eqnarray}
{\cal D}_2\left[F\right](\zeta,\varphi)&\equiv&
\left[\frac{\partial^2}{\partial\zeta^2}+\beta_1 \frac{\partial}{\partial\zeta}+\beta_0\right] F(\zeta,\varphi)
\label{eq:D2a}
\\
&=&\left[
\beta_0-\frac{k_x^2\cos(2\varphi)}{k_0^2}-\sin^2\varphi
-\frac{k_xk_y\sin(2\varphi)}{k_0^2}
\right.\nonumber\\
&&\left.
+\frac{i\beta_1}{k_0}\left(k_x\cos\varphi+k_y\sin\varphi\right)
\right]
F(\zeta,\varphi)
\label{eq:diffint}
\end{eqnarray}
\label{eq:D2}
\end{subequations}
\!\!where $\beta_0$ and $\beta_1$ are $\varphi$-dependent coefficients, and (\ref{eq:diffint}) follows from straightforward differentiation under the integral sign in (\ref{eq:Fint}). Particularly expedient is to choose the coefficients as
\beq
\beta_0=\sin^2\varphi-\frac{\kappa^2}{k_0^2},~~
\beta_1=-\frac{2i\kappa\cos\varphi}{k_0},
\label{eq:betas}
\eeq
which allows recasting (\ref{eq:D2}) in a simplified form
\begin{subequations}
\begin{eqnarray}
{\cal D}_2\left[F\right](\zeta,\varphi)&=&
\frac{\left[\kappa-k_x\cos(2\varphi)-k_y\sin(2\varphi)\right]\left(k_x-\kappa\right)}{k_0^2}
F(\zeta,\varphi)\\
&=&\frac{\kappa\left[\kappa-k_x\cos(2\varphi)-k_y\sin(2\varphi)\right]}{k_0^2}\nonumber\\
&\times&\int\limits_{-\infty }^{\infty}
\frac{\exp\left[i\zeta\left(\displaystyle{\frac{k_x}{k_0}}\cos\varphi+\displaystyle{\frac{k_y}{k_0}}\sin\varphi\right)\right]}{k_x} dk_y.
\end{eqnarray}
\label{eq:DD2}
\end{subequations}
Recalling the spectral-integral representation of the zeroth-order Hankel function of the first kind [cf. (\ref{eq:spectral})],
\beq
\mbox{H}_0^{(1)}(\zeta)=
\int\limits_{-\infty }^{\infty}
\frac{\exp\left[i\zeta\left(\displaystyle{\frac{k_x}{k_0}}\cos\varphi+\displaystyle{\frac{k_y}{k_0}}\sin\varphi\right)\right]}{\pi k_x}dk_y,
\label{eq:H01}
\eeq
the right hand side of (\ref{eq:DD2}) can be readily calculated, yielding
\beq
{\cal D}_2\left[F\right](\zeta,\varphi)=
\left(\frac{\pi\kappa^2}{k_0^2}
\right)
\mbox{H}_0^{(1)}(\zeta)+
\left(
\frac{i\pi\kappa\cos\varphi}{k_0}\right)
\frac{d \mbox{H}_0^{(1)}(\zeta)}{d\zeta}\equiv F_0(\zeta,\varphi).
\label{eq:DE}
\eeq
We have thus demonstrated that the canonical integral in (\ref{eq:Fint}) may be alternatively calculated by solving the second-order, inhomogeneous differential equation in (\ref{eq:DE}) [with ${\cal D}_2$ given in (\ref{eq:D2a}) with (\ref{eq:betas})]. This can be accomplished in a systematic fashion by applying the method of variation of parameters, \cite{Coddington} yielding  
\beq
F(\zeta,\varphi)=A^+ \exp\left(a^+\zeta\right)+
A^- \exp\left(a^-\zeta\right)+Q^+(\zeta,\varphi)+Q^-(\zeta,\varphi),
\label{eq:Fsol}
\eeq
where $A^+$ and $A^-$ are integration constants to be determined (see Appendix \ref{AppD} below), 
\beq
a^{\pm}=-\frac{\beta_1}{2}\pm\sqrt{\frac{\beta_1^2}{4}-\beta_0}
\eeq
are the roots of the characteristic equation, and
\begin{subequations}
\begin{eqnarray}
Q^{\pm}(\zeta,\varphi)&=&\frac{1}{\left(a^{\pm}-a^{\mp}\right)}
\int_{\eta^{\pm}}^{\zeta}
\exp\left[
a^{\pm}\left(
\zeta-\tau
\right)
\right] F_0(\tau,\varphi)d\tau\\
&=&
\frac{i\pi\kappa\cos\varphi}{k_0\left(a^{\pm}-a^{\mp}\right)}\mbox{H}_0^{(1)}(\zeta)\nonumber\\
&+&
\left[
\frac{\pi\kappa\left(
\kappa+ik_0a^{\pm}\cos\varphi
\right)
}{k_0^2\left(a^{\pm}-a^{\mp}\right)}
\right]
\exp\left(a^{\pm}\zeta\right)
{\cal H}\mbox{e}_0^{(1)}\left(
a^{\pm},\eta^{\pm},\zeta
\right)
\label{eq:QQ1}
\end{eqnarray}
\end{subequations}
represents the particular solution,
with the second equality following from (\ref{eq:DE}) [recalling (\ref{eq:CILHI})].

The general solution in (\ref{eq:GenSol1})--(\ref{eq:GenSol4}) immediately follows, by recalling (\ref{eq:recast}) and rearranging terms.\cite{endnote1}

\section{Calculation of the Integration Constants in (\ref{eq:IC1})}
\label{AppD}
The $\varphi$-dependent integration constants $A_{1,2}^{\pm}$ in (\ref{eq:IC1}) can be determined by enforcing the proper boundary conditions (usually, at $\zeta=0$ and for $\zeta\rightarrow\infty$). The general calculation procedure is rather involved, and depends on the sign of the real parts of the $a_{1,2}^{\pm}$ parameters in (\ref{eq:apar}). For instance, if these real parts are all negative, the conditions for $\zeta\rightarrow\infty$ cannot be exploited, and one is led to enforce only the boundary conditions at $\zeta=0$. However, the calculation becomes particularly cumbersome, as the functions $F_{1,2}$ in (\ref{eq:F12}) are nondifferentiable at $\zeta=0$. Nevertheless, for the near-field scenario of interest here, the calculation becomes particularly simple in the two principal planes of the image, namely $x=d$ and $y=d\tan\alpha$, which are the most interesting in order to assess the transverse and longitudinal localization.
 
More specifically, assuming $x_s=0$ and $x=d$, we note that, from (\ref{eq:apar}),
\beq
\mbox{Re}\left(a_{1,2}^-\right)= -\mbox{Re}\left(a_{1,2}^{+}\right)> 0,
\eeq
Accordingly, (\ref{eq:Am}) readily follows by enforcing the regularity-at-infinity conditions 
\beq
\lim_{\zeta\rightarrow\infty} F_{1,2}(\zeta,\varphi)=0.
\eeq
The remaining integration constants are computed by enforcing the conditions at $\zeta=0$, viz.,  
\begin{subequations}
\begin{eqnarray}
F_{1,2}(0,\varphi)&=&
A^+_{1,2}+\chi_{1,2}^+{\cal H}\mbox{e}_0^{(1)}\left(
a_{1,2}^+,\eta_{1,2}^+,0\right)+
\chi_{1,2}^-{\cal H}\mbox{e}_0^{(1)}\left(
a_{1,2}^-,\eta_{1,2}^-,0\right)\\
&=&\kappa_{1,2}\int\limits_{-\infty}^{\infty}\frac{1}{k_x(k_x-\kappa_{1,2})}dk_y=C_{1,2},
\end{eqnarray}
\label{eq:S1}
\end{subequations}
\!\!where the second equality follows from (\ref{eq:Fint}), and the arising spectral integral admits the closed-form analytical solution given in (\ref{eq:C12}). The above derivations also hold approximately for source positions very close to the input slab interface ($x_s\ll \lambda_0$). 

For observations along orthogonal planes passing through the fiducial image position ($y_d=0$, i.e., $y=\tan\alpha$), we note from (\ref{eq:apar}) that
\beq
a_{1,2}^-= a_{1,2}^{+},
\eeq
which basically means that there are only two effective integration constants to be determined. Accordingly, if $\mbox{Re}\left(a_{1,2}^{\pm}\right)< 0$, we can arbitrarily set $A_{1,2}^-=0$ and proceed as above [cf. (\ref{eq:S1})], so that the results in (\ref{eq:IC1}) hold for this case too. Otherwise, if $\mbox{Re}\left(a_{1,2}^{\pm}\right)\ge0$, then the results in (\ref{eq:A120}) readily follow from the regularity-at-infinity conditions.

\newpage

%

%

%

%

%

%

%

%

%


\begin{thebibliography}{99}
\bibitem{endnote1}{Note that the $\mbox{H}_0^{(1)}$ terms in (\ref{eq:QQ1}) cancel out when substituted into (\ref{eq:Fsol}).}
\end{thebibliography}
\end{document}